# Optically induced superconductivity in striped La$_{2-x}$Ba$_x$CuO$_4$ by polarization-selective excitation in the near infrared

D. Nicoletti[1], E. Casandruc[1], Y. Laplace[1], V. Khanna[1,3,4], C. R. Hunt[1,5], S. Kaiser[1],

S. S. Dhesi[3], G. D. Gu[2], J. P. Hill[2], and A. Cavalleri[1,4]

[1] *Max Planck Institute for the Structure and Dynamics of Matter, Hamburg, Germany*

[2] *Condensed Matter Physics and Materials Science Department,*
*Brookhaven National Laboratory, Upton, NY, United States*

[3] *Diamond Light Source, Chilton, Didcot, Oxfordshire, United Kingdom*

[4] *Department of Physics, Clarendon Laboratory, University of Oxford, Oxford, United Kingdom*

[5] *Department of Physics, University of Illinois at Urbana-Champaign, Urbana, Illinois, United States*

**Abstract**

We show that superconducting interlayer coupling, which coexists with and is depressed by stripe order in La$_{1.885}$Ba$_{0.115}$CuO$_4$, can be enhanced by excitation with near-infrared laser pulses. For temperatures lower than $T_c$ = 13 K, we observe a blue-shift of the equilibrium Josephson plasma resonance, detected by terahertz-frequency reflectivity measurements. Key to this measurement is the ability to probe the optical properties at frequencies as low as 150 GHz, detecting the weak interlayer coupling strengths. For $T > T_c$ a similar plasma resonance, absent at equilibrium, is induced up to the spin-ordering temperature $T_{SO} \simeq 40$ K. These effects are reminiscent but qualitatively different from the light-induced superconductivity observed by resonant phonon excitation in La$_{1.675}$Eu$_{0.2}$Sr$_{0.125}$CuO$_{6.5}$. Importantly, enhancement of the below-$T_c$ interlayer coupling and its appearance above $T_c$ are preferentially achieved when the near-infrared pump light is polarized perpendicular to the superconducting planes, likely due to more effective melting of stripe order and the less effective excitation of quasiparticles from the Cooper pair condensate when compared to in-plane excitation.

Charge and spin orders in cuprates[1,2,3,4,5] and their interaction with the superconducting state are believed to harbor important clues to the mechanism of high-temperature superconductivity. A prototypical case is that of *stripes* in the $CuO_2$ planes of single-layer La-based compounds. These consist of one-dimensional chains of doped holes separating regions of oppositely phased antiferromagnetism.

Here, we study $La_{2-x}Ba_xCuO_4$ (LBCO), a single-layer compound that displays complete suppression of the transition temperature $T_c$ at $x = 1/8$ (Ref. 6, 7) and a coexistence between stripes and superconductivity at lower doping values (see Fig. 1(a)). Stripe order in LBCO is stabilized by the so-called low-temperature tetragonal (LTT) phase[8,9], a structural phase in which the fourfold rotational symmetry of the $CuO_2$ planes is broken by the buckling of the oxygen octahedra in alternate planes[10].

Yet, stripes have also been observed without any LTT distortion[11,12], underscoring the notion that charge ordering alone may be driving the microscopic physics. Recent experiments[13] suggest that the individual striped planes may in fact be in a highly coherent paired state, a so-called *pair density wave*, in which the charge stripes (see Fig. 1(b)) modulate the superconducting order parameter in plane and frustrate the interlayer coherent transport[14].

Ultrafast optical experiments have brought new evidence in support of this view. Excitation of the in-plane Cu-O stretching mode in non-superconducting, LTT distorted[15] $La_{1.675}Eu_{0.2}Sr_{0.125}CuO_4$ ($LESCO_{1/8}$)[16] was shown to promptly establish superconducting interlayer tunneling[17,18], suggesting ultrafast removal of frustration. Under similar excitation conditions, femtosecond resonant soft x-ray diffraction showed that charge and lattice order in 1/8-doped LBCO are decoupled on ultrafast timescales[19], further indicating that charge modulation alone, and not the LTT distortion, may be preventing interlayer coupling at equilibrium.

Here, we seek alternative means to remove stripe order in LBCO. Rather than driving the lattice in the mid infrared, we use high-energy (1.5 eV) photo-excitation, which has often been applied to melt charge density waves[20,21,22,23,24,25,26]. Despite the significant excess energy introduced, we show that for pump polarization perpendicular to the Cu-O planes one can enhance the interlayer coupling. We find that this effect can be achieved preferentially at doping levels for which the stripes coexist with superconductivity at low temperatures, with the effect being strongest in $La_{1.885}Ba_{0.115}CuO_4$.

The $La_{2-x}Ba_xCuO_4$ single crystals used in our experiments were grown[7] at three nominal Ba concentrations $x$ = 9.5%, 11.5%, and 12.5%, covering the most underdoped region of the phase diagram. The $x$ = 9.5% and 11.5% compounds are superconducting with transitions at $T_c \simeq$ 32 and 13 K, respectively. At 9.5% doping, charge- and spin-order, the structural transition and superconductivity all appear at the same temperature $T_{CO} \simeq T_{SO} \simeq T_{LT} \simeq T_c \simeq$ 32 K. These transitions become decoupled at higher hole concentrations, as $T_{CO} \simeq T_{LT} \simeq$ 53 K and $T_{SO} \simeq$ 41 K at $x$ = 11.5%. For the commensurate doping $x$ = 1/8, $T_{CO} \simeq T_{LT} \simeq$ 55 K and $T_{SO} \simeq$ 42 K, with superconductivity being suppressed to below the lowest measurable temperatures, $T_c \lesssim$ 2.4 K[7,27].

The crystals were cut and polished to give an *ac* surface with large enough area (between 5 and 15 mm$^2$) to perform long-wavelength THz spectroscopy. The equilibrium optical properties at all three doping levels were determined using single-cycle THz pulses generated by illuminating a photoconductive antenna with near-infrared laser pulses from a Ti:Sa amplifier. These probe pulses were focused onto the sample surface, with polarization perpendicular to the $CuO_2$ planes. The reflected electric field was measured by electro-optic sampling in ZnTe at different temperatures, both below and above $T_c$ (see circles on phase diagram in Fig. 1(a)). By making use of the broadband reflectivity from Ref. 27, measured in the same batch of samples, all the equilibrium complex optical functions were determined[28].

The *c*-axis equilibrium reflectivities are shown in the left-hand panel of Fig. 1(c) at $T$ = 5 K. For both $x$ = 9.5% and 11.5% a Josephson plasma resonance (JPR) appears[29] below $T_c$ at ∼ 500 GHz ($x$ = 9.5%) and at ∼ 200 GHz ($x$ = 11.5%). Note that the antennas used in our experiments generated sufficient spectral weight between 150 and 3000 GHz. The 150-200 GHz frequency range (∼ 5-7 cm$^{-1}$) is difficult to reach with conventional infrared spectroscopy, and to our knowledge, the edge reported here is the first observation of a static JPR in $La_{1.885}Ba_{0.115}CuO_4$. The THz reflectivity of the $x$ = 12.5% sample is featureless, with no evidence of a JPR[27].

The LBCO crystals were then photo-excited with ∼100 fs, 800-nm wavelength laser pulses, polarized along the *c* axis. These pulses were focused on a ≳2-mm-diameter spot, at ∼2 mJ/cm$^2$ fluence. The pump photon energy is indicated by arrows in the equilibrium reflectivity and optical conductivity spectra (Fig. 1(c)). Note that the polarization of our pump pulses makes our experiments different from previous studies performed to date in cuprates, for which charge excitation was typically reported with pump fields polarized in the $CuO_2$ planes[30,31,32,33,34,35,36,37,38,39].

The complex *c*-axis optical properties could be retrieved as a function of pump-probe delay with a temporal resolution set by the inverse bandwidth of the THz probe pulse[17,18,40], *i.e.* $\Delta\tau \simeq 350$ fs. The mismatch between the penetration depth of the near-infrared pump (about 400 nm) and of the THz probe (∼ 50-500 μm) was taken into account by modeling the response of the system as that of a thin photo-excited layer on top of an unperturbed bulk[41].

The recalculated reflectivity $R(\omega)$ of the top photo-excited LBCO layer is shown in Figure 2 for all measured doping values and temperatures. In the figure, the spectra at +1.5 ps pump-probe delay are displayed alongside the same quantity at equilibrium (gray). At the lowest doping ($x$ = 9.5%) and temperature ($T < T_c$), a small blue-shift of the JPR is observed (panel a.1), suggestive of a light-induced increase in the interlayer Josephson coupling. For the same material above $T_c$ (panel a.2), no significant photo-induced dynamics could be measured.

The response of the $x$ = 11.5% sample is far more striking. Below $T_c$ (panel b.1) a prompt blue-shift of the equilibrium JPR from ~200 to 600 GHz was observed. Even more remarkably, in the spin-ordered phase (panel b.2) above $T_c$, where no edge is observed at equilibrium, an edge appeared at ~500 GHz immediately after photo-excitation. A similar effect, although less evident, was present also above $T_{SO}$ (panel b.3), where a reflectivity edge appeared near ~400 GHz. No appreciable photo-induced dynamics could be measured above $T_{CO}$ (panel b.4). Finally, in the 1/8-doped material no effect was detected in our measured frequency range at any temperature and pump-probe delay (panels c).

In the following we restrict our analysis to the $x$ = 11.5% compound, where the effect was strongest. The complete time-dependent response of the interlayer plasmon is reported by plotting the energy loss function $\left[-\operatorname{Im}(1/\tilde{\varepsilon})\right]$ in the color plots of Figure 3. Selected line cuts are shown in the upper panels.

Below $T_c$ the loss-function peak was found to continuously shift from its equilibrium value (~200 GHz) toward higher frequencies, up to ~1200 GHz (at ~2.5 ps delay) before relaxing back to lower values. Also, for $T_c < T < T_{SO}$ (panels b) a well-developed loss-function peak (absent at equilibrium), appeared in the perturbed material, continuously broadening and shifting first to the blue and then to the red at longer time delays. As discussed above, all light-induced effects progressively reduce with increasing $T$ and completely disappear after crossing $T_{CO}$ (panels c and d).

Further analysis is reported in Figure 4, where the transient optical conductivity $\sigma_1(\omega)+i\sigma_2(\omega)$ is displayed for $T$ = 5 K (panels a.1-a.2), and $T$ = 30 K (panels b.1-b.2) at three selected pump-probe delays: $\tau < 0$ (equilibrium), $\tau$ = 1.5 ps, and $\tau$ = 5 ps. In the superconducting state at equilibrium, a fully gapped $\sigma_1(\omega)$ (Fig. 4(a.1), gray curve) and a $\sigma_2(\omega)$ that turn positive and increases with decreasing $\omega$ is shown (Fig. 4(a.2), gray curve). At about 1.5 ps after photo-excitation (red curves), a strong enhancement in $\sigma_2(\omega)$ was observed down to the lowest measured frequency, while $\sigma_1(\omega)$ remained gapped. This complex conductivity behavior underscores the increase in the superconducting coupling between layers and is incompatible with a charge excitation scenario. Indeed, the light-induced conductivity changes saturate with fluence (at ~1 mJ/cm$^2$) and do not follow the response of an incoherent plasma excited above a semiconducting gap. At later delays ($\tau$ = 5 ps, blue curves), a relaxation towards a more incoherent state is observed (see later discussion).

The same qualitative behavior was found at 30 K, where the system is insulating at equilibrium. After photo-excitation, $\sigma_2(\omega)$ shows a strong light-induced enhancement at low frequency, turning positive and increasing with decreasing $\omega$ down to the lowest measured frequency, strongly resembling the response observed below $T_c$. These transient properties could be quantified by fitting the optical response with a Drude model:

$$\sigma_1(\omega) + i\sigma_2(\omega) = \frac{\omega_P^2}{4\pi} \frac{\tau_S}{1 - i\omega\tau_S}$$

where $\omega_P$ is the plasma frequency and $\tau_S$ the carrier scattering time. In order to mimic the background in the optical spectra, mainly caused by phonon absorptions (see Fig.

1(c)), high-frequency Lorentz oscillators have been added, chosen for the equilibrium compound and kept constant in all fits. Examples of these fits are reported as black dots in Fig. 4 (b.1-2). Remarkably, the experimental data could be well reproduced at all temperatures and time delays. The extracted fit parameters, *i.e.* the screened plasma frequency $\widetilde{\omega}_P = \omega_P/\sqrt{\varepsilon_{FIR}}$ (where $\varepsilon_{FIR} \simeq 30$, see Ref. 27) and the scattering time $\tau_S$ are displayed in Fig. 4 (a.3-4) and (b.3-4).

Below $T_c$ (panels a.3-4) and at delays $\tau \lesssim 1.5$ ps the transient optical properties are described by a Drude model with scattering time $\tau_S > 5$ ps, or with the spectrum of a superconductor with $\tau_S \to \infty$:

$$\sigma_1(\omega) + i\sigma_2(\omega) = \frac{\omega_P^2}{8}\delta[\omega = 0] + \frac{\omega_P^2}{4\pi}\frac{i}{\omega}.$$

Similar dynamics can be extracted also from the fits to the 30 K data (panels b.3-4) where, from the insulating ground state (gray region), a state with high-mobility carriers ($\tau_S > 5$ ps) is induced, whose optical properties are compatible with those of a transient superconductor.

At longer time delays ($\tau \gtrsim 2$ ps), the system relaxes into a state only quantitatively different, in which coherence is reduced. Here the data can be fitted with a finite carrier scattering time $\tau_S \sim 5$ ps, reducing to $\tau_S \sim 1$ ps at later delays. At the same time, $\widetilde{\omega}_P$ continues to increase, exceeding 1 THz at $\tau \simeq 2.5$ ps, and then relaxing back to about 500 GHz.

In this relaxed state the scattering times are still anomalously high for conventional incoherent charge transport, and rather could be interpreted as indicative of a coherent state in which fluctuations are important, in analogy with interpretation of the equilibrium THz response of $La_{2-x}Sr_xCuO_4$ below and above $T_c$ (Ref. 42).

The polarization-selective character of the enhanced coherent response is highlighted in Figure 5, where the transient optical conductivity of Figure 4(b) at 30 K (red) is compared with that obtained after excitation with light polarized in the CuO$_2$ planes (black). Not only is the photo-induced interlayer coupling, evidenced by the low-$\omega$ increase in $\sigma_2(\omega)$, far more evident after excitation perpendicular to the planes, but the quasiparticle response in $\sigma_1(\omega)$ is very weak in this case and larger for in-plane excitation.

The idea that interlayer coherent coupling is preferentially enhanced when light is polarized perpendicular to the planes should be further investigated by a quantitative theoretical analysis. However, qualitative considerations can aid us at this stage. Indeed, light polarized out-of-plane does not couple to quasiparticle excitations in two-dimensional superconductors, and is expected to couple only weakly in the quasi-two-dimensional case of cuprates. On the other hand, because of the peculiar arrangement of charge order in LBCO, in which parallel stripes in planes 1 and 3 (and 2 and 4) are shifted by $\pi$ (see Fig. 1(b)), coupling to the charge stripes and dipole activity is expected for this polarization[43]. Indeed, because optical melting of stripe order is a promising and not yet optimized means to affect this competition, future studies should thoroughly analyze its dependence on the excitation wavelength.

In summary, enhancement of Josephson coupling was achieved by optical excitation with light polarized perpendicular to the CuO$_2$ planes of LBCO. The data is interpreted by assuming that such optical perturbation melts spin and charge order, which competes with interlayer superconducting transport at equilibrium. The absence of any light-induced response in the 1/8-doped sample may be attributed to a lower photo-susceptibility of the stripes in this compound (which has the largest stripe order parameter and correlation length at equilibrium[7]) or because a light-induced

JPR occurs at frequencies below our probing window (< 150 GHz). Further investigations with time-resolved x-ray diffraction are required to demonstrate that such effect, observed in the optical response at THz frequencies, is related to a light-induced melting of the competing stripe phase, and to determine the fate, if any of the LTT distortion.

**Acknowledgments**

The research leading to these results has received funding from the European Research Council under the European Union's Seventh Framework Programme (FP7/2007-2013)/ERC Grant Agreement No. 319286 (Q-MAC), and from the German Research Foundation (DFG-SFB 925). Work performed at Brookhaven was supported by US Department of Energy, Division of Materials Science under contract no. DE-AC02-98CH10886.

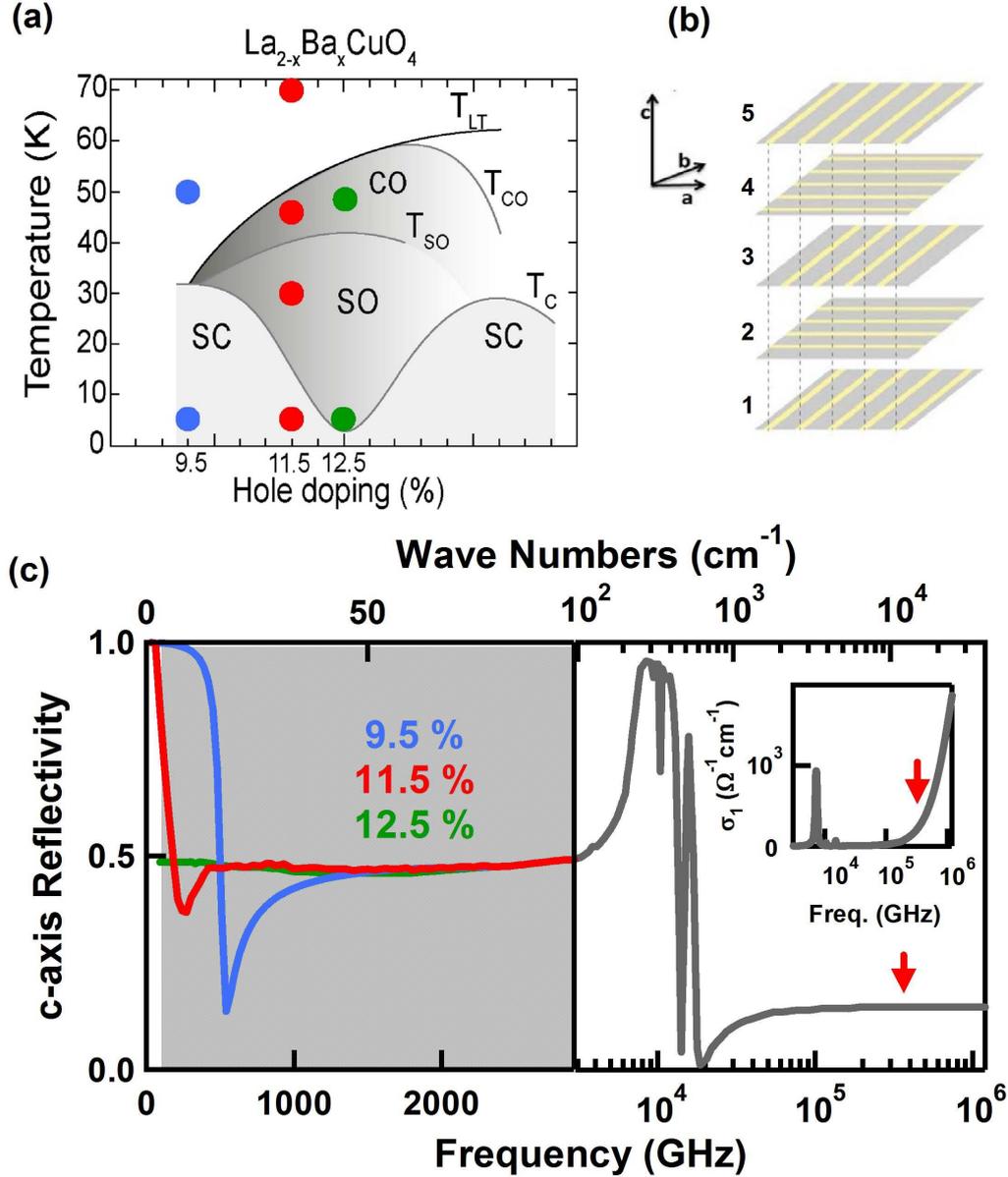

**Figure 1.** (a) Temperature-doping phase diagram of LBCO, as determined in Ref. 7. $T_c$, $T_{CO}$, $T_{SO}$, and $T_{LT}$ indicate the superconducting, charge-order, spin-order, and structural transition temperatures, respectively. Colored circles indicate the different dopings and temperatures for which data are reported here. (b) Periodic stacking of $CuO_2$ planes in the stripe phase. The stripe orientation rotates by 90° between layers. (c) Equilibrium $c$-axis optical properties of LBCO. Left panel: THz reflectivity of the three samples at $T$ = 5 K. The region investigated in this experiment is shaded in gray. Right panel and inset: broadband $c$-axis reflectivity and optical conductivity of LBCO from Ref. 27. Red arrows indicate the pump photon energy.

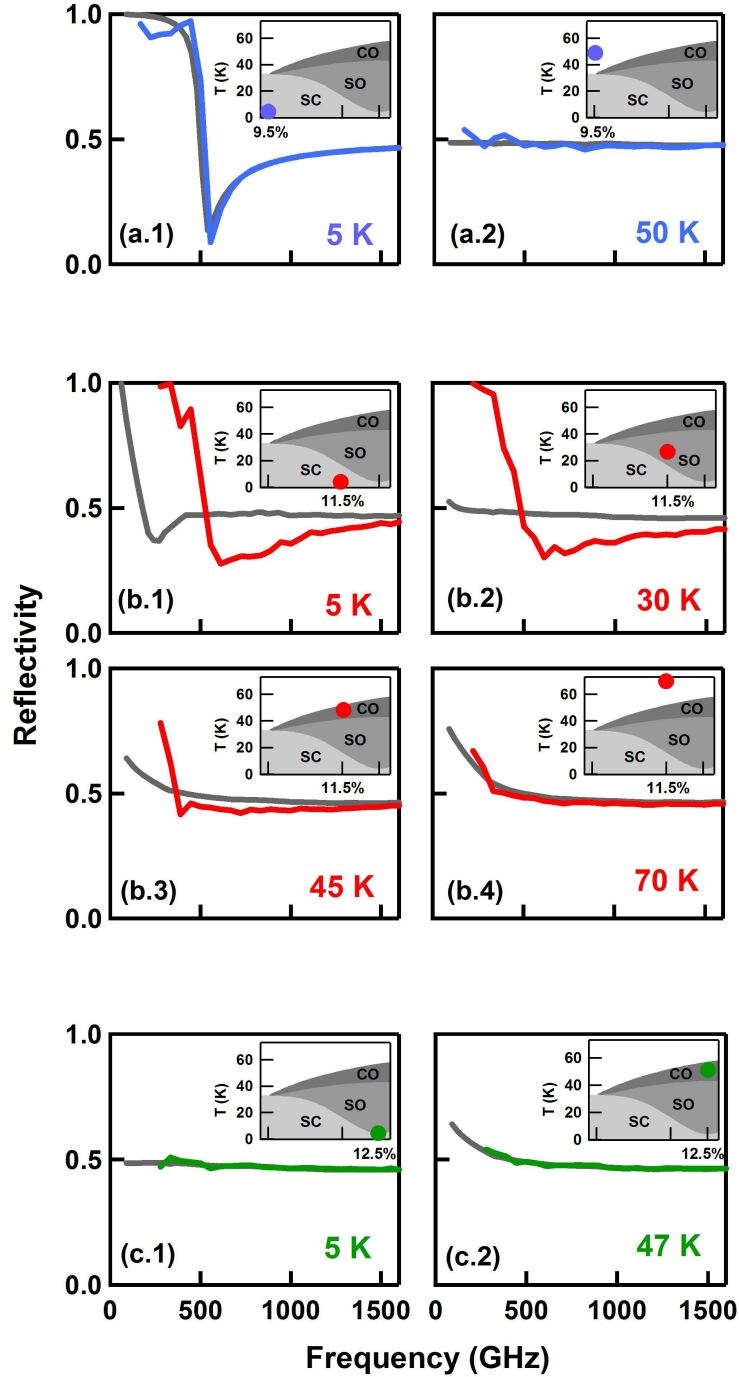

**Figure 2.** THz reflectivity of LBCO displayed at different doping values and temperatures at equilibrium (gray) and 1.5 ps after excitation (colored). Data in panels a and b have been taken with a pump fluence of ∼2 mJ/cm$^2$ (a saturation in the fluence dependence of the pump-induced changes was found above ∼1 mJ/cm$^2$). Those in panels c were taken instead with ∼3 mJ/cm$^2$. All spectra are shown in the region below 1600 GHz to highlight the pump-induced changes. The low-frequency cutoff is due to sample-size effects. Transient spectra cover a narrower range, due to the higher noise of the pump-probe measurement.

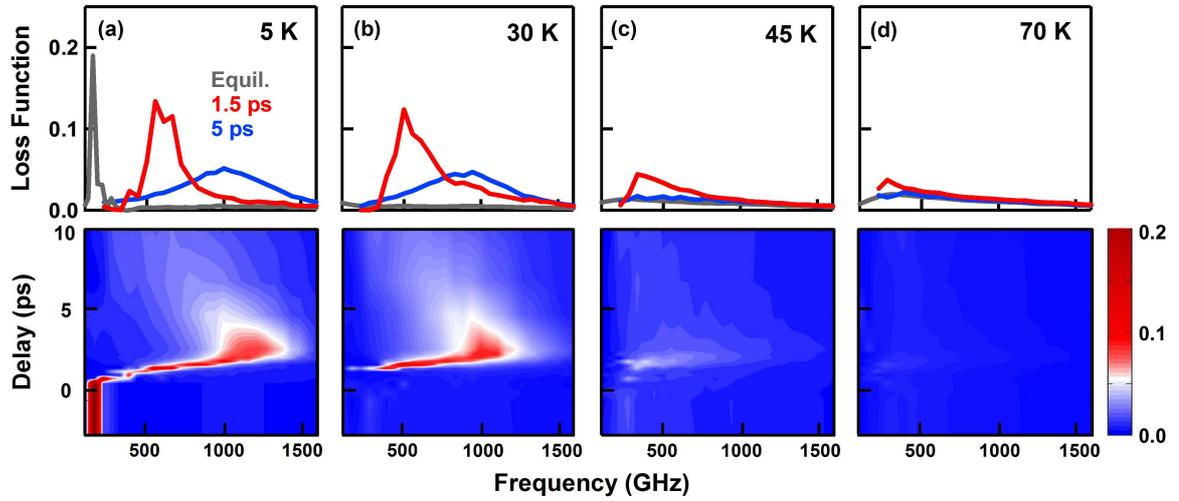

**Figure 3.** Energy loss function $[-\mathrm{Im}(1/\tilde{\varepsilon})]$ of $La_{1.885}Ba_{0.115}CuO_4$ as a function of temperature and pump-probe delay. The lower panels show its evolution throughout the light-induced dynamics. The upper panels show selected line cuts at negative (gray), +1.5 ps (red), and +5 ps (blue) time delay. All data have been taken using a pump fluence of 2 mJ/cm$^2$.

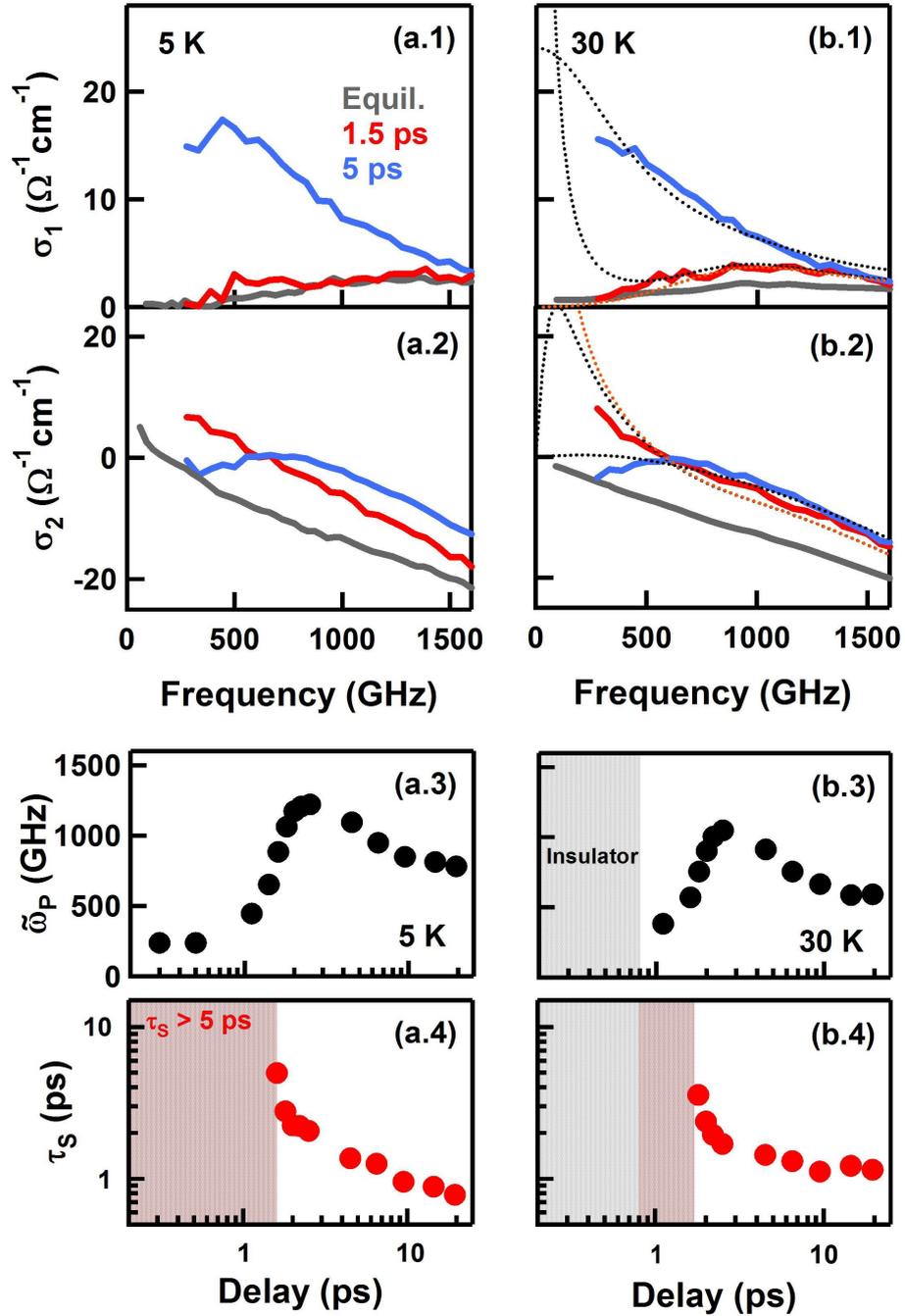

**Figure 4.** Panels a.1-2 and b.1-2: Complex optical conductivity of La$_{1.885}$Ba$_{0.115}$CuO$_4$ at 5 K and 30 K, shown at different pump-probe delays, for a fluence of 2 mJ/cm$^2$. In panels b.1-2 examples of fits with a Drude model (black dots) and with a perfect-conductor model (red dots) are displayed. Panels a.3-4 and b.3-4: Parameters extracted from the Drude fits as a function of pump-probe delay. The gray shaded region indicates the insulating regime (no Drude fit possible). The red shaded area refers to the highly coherent state (see main text).

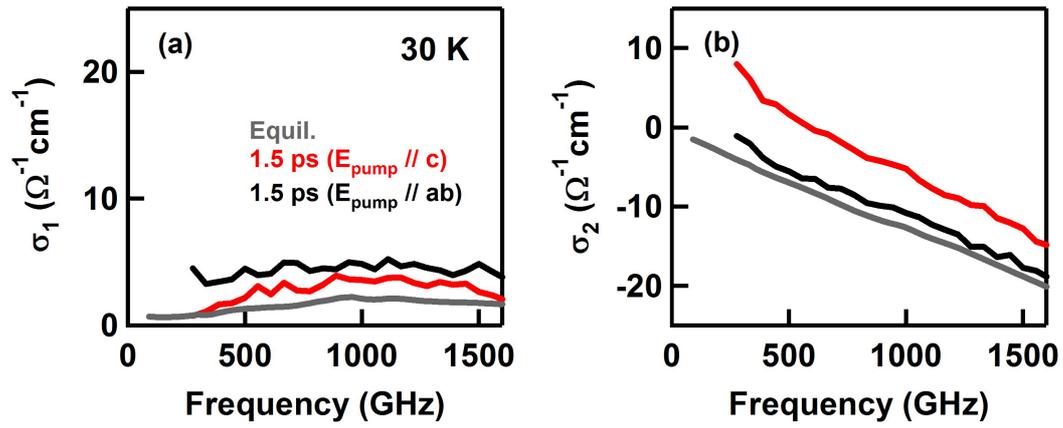

**Figure 5.** Complex optical conductivity of $La_{1.885}Ba_{0.115}CuO_4$ at 30 K, 1.5 ps after optical excitation with light polarized perpendicular (red) or parallel (black) to the $CuO_2$ planes. Data are taken with the same pump fluence of 2 mJ/cm².

[41] The pump-induced changes in the amplitude and phase of the reflected THz electric field were measured at different pump-probe delays. These "raw" reflectivity changes were only ~0.5-1% due to the pump-probe penetration depth mismatch. At THz frequencies, the probe interrogates a volume that is between $10^2$ and $10^3$ times larger than the transformed region beneath the surface, with this mismatch being a function of frequency. Such mismatch was taken into account by modeling the response of the system as that of a homogeneously photo-excited thin layer on top of an unperturbed bulk (which retains the optical properties of the sample at equilibrium). By calculating the coupled Fresnel equations of such multi-layer system [M. Dressel and G. Grüner, Electrodynamics of Solids, Cambridge University Press, Cambridge (2002)], the transient optical response (reflectivity, energy loss function, complex optical conductivity) of the photo-excited layer could be derived. A detailed explanation of this procedure can also be found in Ref. 18. The calculated optical properties were then compared with those obtained by treating the excited surface as a stack of thinner layers with a homogeneous

refractive index and describing the excitation profile by an exponential decay (see Ref. 17, 36). The results of the two models were found to be in agreement within a few percent.